# CRYSTAL SIMULATIONS: THE ROAD FROM THE SPS TO THE LHC

V.M. Biryukov [#], Institute for High Energy Physics, 142281 Protvino, Russia
Invited talk at CC-2005: Crystal Collimation in Hadron Storage Rings (CERN, 7-8 March 2005)
http://care-hhh.web.cern.ch/CARE-HHH/CrystalCollimation/

*Abstract*
  The understanding of the crystal collimation and extraction experiments performed in the recent decade at CERN SPS, FNAL Tevatron, IHEP U-70 and RHIC is reviewed from the standpoint of Monte Carlo simulations. The expectations for the LHC - the crystal efficiency and the requirements to a crystal - are outlined basing on the same computer model verified in the above experiments. Finally, we discuss key issues for the future experiments at CERN SPS and Tevatron aiming to reach the most efficient channeling and to bring these tests to the future LHC as close as possible.

## INTRODUCTION

The idea to deflect proton beams using bent crystals, originally proposed by E.N. Tsyganov in 1976 [1] and demonstrated in 1979 in Dubna [2] on proton beams of a few GeV energy has received strong development. The physics related to channeling mechanisms was studied in details, first in the early 1980's at St.Petersburg, Dubna, CERN, and FNAL using proton beams of 1 to 800 GeV (see Refs., e.g. in [3]). Crystal-assisted extraction from accelerator was demonstrated for the first time in 1984 in Dubna at beam energies of 4-8 GeV [4].

Crystal extraction and collimation experiments have greatly progressed in recent decade, spanning over three decades in energy from 1 GeV [5] through nearly 1 TeV [6-8]. The theory of crystal extraction is essentially based on Monte Carlo simulations [9-13], as the extraction process includes multiple passes through the crystal, and turns in the accelerator, of the beam particles.

Even more importantly, tracking of a particle through a bent crystal lattice requires not only a calculation of a particle dynamics in this nonlinear field, but also a generation of random events of scattering on the crystal electrons and nuclei.

To track particles through the curved crystal lattices in simulation we apply the approach with a continuous potential introduced by Lindhard [14]. In this approach one considers collisions of the incoming particle with the atomic strings or planes instead of with separate atoms, if the particle is sufficiently aligned with respect to the crystallographic axis or plane.

 The typical step size along the crystal length in simulation is about 1 micron or a fraction of it, as defined by the particle dynamics in crystal channel. By every step the probabilities of scattering events on electrons and nuclei are computed depending on their local densities, which are functions of coordinates.

This ensures correct orientational dependence of all the processes in crystal material. Further details on the simulation code may be found in Refs.[10-13].

Leaving aside the details of channeling physics, it may be useful to mention that accelerator physicist will find many familiar things there:

- Channeled particle oscillates in a transverse nonlinear field of a crystal channel, which is the same thing as the *betatron oscillations* in accelerator, but on a much different scale (the wavelength is 0.1 mm at 1 TeV in silicon crystal). The number of oscillations per crystal length can be several thousand in practice. The concepts of beam emittance, or particle action have analogue in crystal channeling.
- The crystal nuclei arranged in crystallographic planes represent the "*vacuum chamber walls*". Any particle approached the nuclei is rapidly lost from channeling state. Notice a different scale again: the "vacuum chamber" size is ~2 Angstroms.
- The well-channeled particles are confined far from nuclei (from "aperture"). They are lost then only due to scattering on electrons. This is analogy to "*scattering on residual gas*". This may result in a gradual increase of the particle amplitude or just a catastrophic loss in a single scattering event.
- Like the real accelerator lattice may suffer from *errors of alignment*, the lattice of real crystal may have dislocations too, causing an extra diffusion of particle amplitude or (more likely) a catastrophic loss [13].
- Accelerators tend to use low temperature, superconducting magnets. Interestingly, the crystals cooled to *cryogenic temperatures* are more efficient, too [15].

## THE SPS EXPERIMENTS

A detailed account for the crystal extraction experiments made at the CERN SPS can be found in this volume and in refs. [16-18].

Before these SPS studies, the theoretical comparisons [9] with extraction experiments [4,19] were restricted by analytical estimates only, which gave the right order of magnitude. The computer simulations considered idealized models only and predicted the extraction efficiencies always in the order of 90-99% (see e.g. [9,20]) while real experiments handled much smaller efficiencies, in the order of 0.01% [4,19].

The considered-below theoretical work has been the first and rather detailed comparison between the realistic calculation from the first principles (computer simulation) and the experiment.

The simulation was performed [21-22] with parameters matching those of the SPS experiment. Over $10^5$ protons have been tracked both in the crystal and in the accelerator for many subsequent passes and turns until they were lost either at the aperture or in interaction with crystal nuclei.

In the simulation, different assumptions about quality of the crystal surface were applied: one was an ideal surface, whereas the other one assumed near-surface irregularities (a `septum width') of a few µm due to a miscut angle (between the Si(110) planes and the crystal face) 200 µrad, surface nonflatness 1 µm, plus 1 µm thick amorphous layer superposed. Two options were considered. The *first*, with impact parameter below 1 µm and surface parameters as described above, excludes the possibility of channeling in the first pass through the crystal. This is compared to the *second* option, in which the crystal surface is assumed perfect, i.e., with a zero septum width.

Table 1 shows the expected extraction efficiencies for both options from the first simulation run and the measured lower limit of extraction efficiency [23,24] at that early time.

Table 1: SPS crystal extraction efficiencies from the early runs, Monte Carlo prediction and experiment.

| Option | Monte Carlo | Experiment |
| --- | --- | --- |
| Poor surface | 15% | greater than 2-3% (lower limit only measured) |
| Ideal surface | 40% | |

Though the efficiency comparison, theory to measurements, was not possible at that time, from the analysis of the simulation results one could see that the perfect-surface simulation predicted narrow high peaks for the angular scans (30 µrad FWHM) and extracted-beam profiles, which have not been observed.

The imperfect-surface option, however, is approximately consistent with the experimental observations: wide (about 200 µrad FWHM) angular scan and sophisticated profiles of the extracted beam (dependent on the crystal alignment).

The efficiency was measured in the SPS experiment with the first tested crystals to be 10±1.7% [16]. The detailed simulations [21] have shown that efficiency should be a function of the vertical coordinate of the beam w.r.t. the crystal (for its given shape), and be from 12 to 18% at peak, with imperfect-surface option.

The simulation studies for a new crystal with another geometry (``U-shaped'') were performed prior [21] to the measurements. The model followed the parameters and design of this crystal, with the same SPS setting.

Again the two options, an imperfect or perfect edge, have been studied. Figure 1 (borrowed from ref. [25]) shows the angular scan (as narrow as 70 µrad FWHM) of crystal extraction predicted in simulations [21] for the U-shaped crystal with edge imperfections in comparison to the measurements [26], in a rather good agreement. The peak efficiency, ~19%, was expected to be slightly increased with the new crystal. For an ideal crystal and a parallel incident beam, the simulation predicted a peak efficiency of ~50% and a very narrow angular scan (25 µrad FWHM).

Another SPS experiment employed a crystal with an amorphous layer at the edge to suppress the channeling in the first passage of the protons [17]. The extraction efficiency with this crystal was indeed of the same order of magnitude as found without an amorphous layer, thus confirming the theoretical prediction [21-22] that the first-pass channeling is suppressed in the SPS crystals.

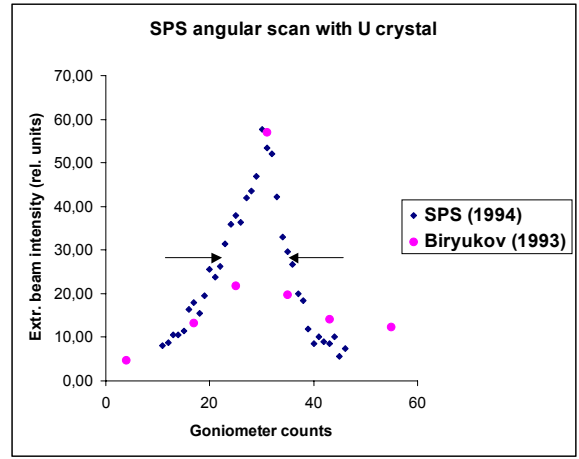

Figure 1. The angular scan of extraction with a U-shaped crystal. Prediction (1993) and measurement (1994).

## THE TEVATRON EXPERIMENT

The Tevatron extraction experiment has provided another check of theory at a much higher energy of 900 GeV. A detailed report of predictions for this experiment from the Monte Carlo simulations was published [27] two years before the measurements were taken [7]. The simulation predicted the efficiency of 35% for a realistic crystal in the Tevatron experiment [27].

During the FNAL test, the halo created by beam-beam interaction in the periphery of the circulating beam was extracted from the beam pipe without measurable effect on the background seen by the experimental detectors. The crystal was channeling a 900-GeV proton beam with an efficiency of ~30% [7], showing a rather good agreement with the theoretical expectation.

Apart from observing the channeled particles, this experiment has measured also the particles dechanneled from the crystal, appearing as a tail. The number of particles in the visible tail was measured 20% of the peak [7]. A simulation of the experiment predicted 25% [27].

## CRYSTAL OPTIMISATION

The length of the Si crystals used in the SPS and Tevatron experiments was about optimal to bend protons with a *single* pass. The efficiency of the *multi*-pass extraction is defined by the processes of channeling, scattering, and nuclear interaction in the crystal, which depend essentially on the crystal length *L*.

In order to let the circulating particles encounter the crystal many times and suffer less scattering and nuclear interactions in the crystal, one has to minimise the crystal length down to some limit set by the physics of channeling in a strongly bent crystal [9,28]. This optimisation was studied in Monte Carlo simulations in general and for the experiments at CERN SPS and the Tevatron [9,21,27,29-31], taking into account the circulation of particle in the accelerator ring over many turns with multiple encounters with a bent crystal.

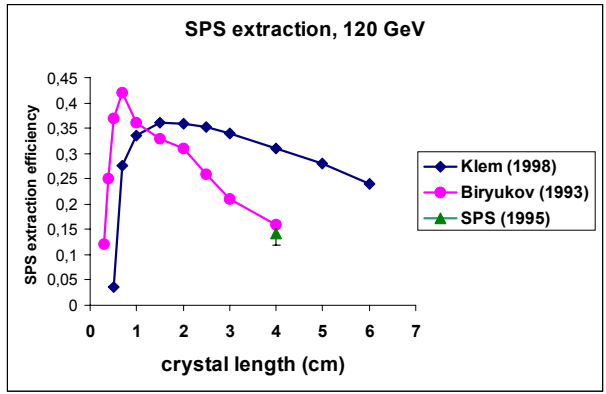

Figure 2. The SPS extraction efficiency plotted vs crystal length. Simulations (1993, 1998) and measurement for the 4-cm U-shaped crystal (1995).

The optimization with the simulations was made with the assumption of a uniform crystal curvature and is shown in Figure 2 for the case of SPS at 120 GeV [21-22]. For a crystal with an imperfect surface there is important dependence on length. A new optimum around $L \approx 0.7$ cm more than doubles the efficiency as compared to that for the 4 cm crystal. Predicting this boost in efficiency was not a trivial matter: Fig. 2 shows also the absolute efficiency from another simulation [32], which predicted just 15% rise from a change of crystal length from 4 to 2 cm. The same Figure shows the SPS measurement with a U-shaped crystal of 4 cm length.

Similarly, for the Tevatron E853 experiment it was found that the efficiency of extraction could again be increased with the use of a shorter crystal. The efficiency is maximal, near 70%, in the crystal length range from 0.4 to 1.0 cm [27,29].

Much of these dependences, and even the absolute figures of efficiency of multi-pass extraction, can be understood in the framework of analytical theory crystal extraction. We refer to [30,31] for discussion of it, together with more simulations repeated for the energies of 14 and 270 GeV, where new measurements have been done at the SPS.

## THE IHEP EXPERIMENT ON CRYSTAL COLLIMATION

Monte Carlo study done for the experiment undertaken at IHEP has predicted [33] that a crystal can be shorten quite a bit, down to ~1 mm along the 70-GeV beam in the extraction set-up of IHEP U-70. However, the benefits from this optimisation were expected tremendous: the crystal extraction efficiency could be as high as over 90%. Figure 3 shows both the predicted [33] dependence of the IHEP crystal extraction efficiency as a function of the crystal length, and some history of the measurements since 1997 [5,34-40].

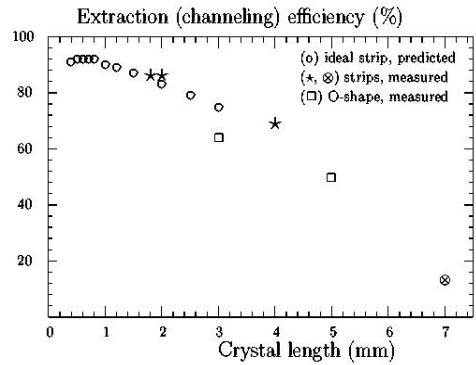

Figure 3. Crystal extraction/collimation efficiency for 70 GeV protons vs crystal length along the beam. IHEP measurements and Monte Carlo prediction.

Producing bent crystal deflectors of required size and curvature is not an easy task, moreover as one takes into account that deflector has to be placed in a circulating beam and any extra disturbance to halo particles must be avoided. It took several years in IHEP to approach the target set by theory. First IHEP crystal was a kind of strip, 7 mm along the 70 GeV beam, tested at the end of 1997. Then IHEP turned to analog of U-shaped crystals of SPS; the required deflectors were cut and polished in the optical workshop of PNPI. The decisive step was invention in IHEP of strip-type deflectors (Fig. 4), very short – down to ~2 mm along the beam, without straight parts and uniformly bent. This breakthrough in bent crystal technology in IHEP has lead to the dramatic boost in crystal

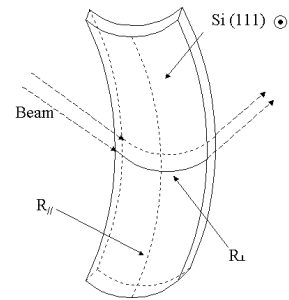

efficiency.
Figure 4. Strip-type bent crystal invented in IHEP.

Many strip deflectors were produced in IHEP from commercially available wafers of different origin. Now crystal systems extract 70 GeV protons from IHEP main ring with efficiency of 85% at intensity of $10^{12}$. Today, six locations on the IHEP 70-GeV main ring are equipped by crystal extraction systems, serving mostly for routine applications rather than for research. The experimentally recorded high efficiency followed nicely the prediction as seen in Figure 3. Compared to the CERN SPS and Tevatron experiments, the efficiency is improved by a factor of 3-8 while the crystal size along the beam was reduced by a factor of 15-20 (from 30-40 mm to less than 2 mm).

Experimentally, the extraction efficiency in IHEP was defined as the ratio of the extracted beam intensity as measured in the external beam line to all the beam loss in the circulating beam. A remarkable feature of the IHEP extraction is that the record high efficiency of about 85% is pertained even when the *entire beam* stored in the ring is dumped onto the crystal.

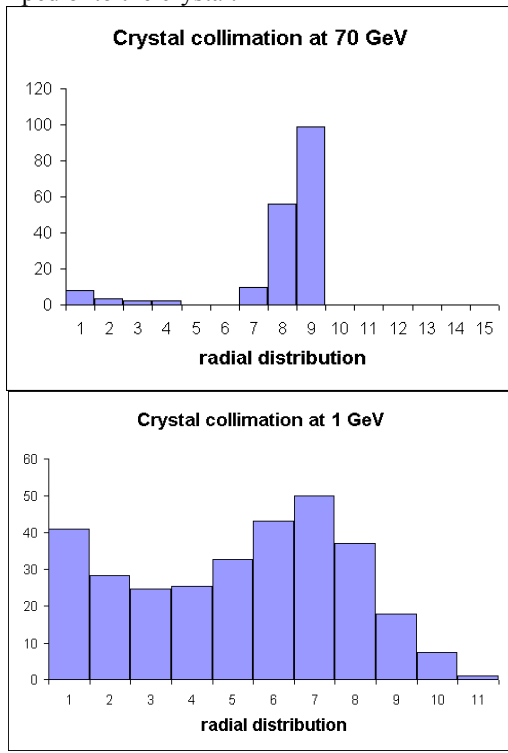

Figure 5. The radial beam profile observed at the entry face of the collimator with crystal working as a primary scraper, at top energy (70 GeV, top) and at injection plateau (1.3 GeV, bottom); crystal is the same.

IHEP has many locations on the U-70 ring where crystals are installed for extraction and collimation studies. Two of these locations are dedicated for crystal collimation. In a collimation experiment, a bent crystal is positioned upstream of a secondary collimator (stainless steel absorber 4 cm wide, 18 cm high, 250 cm long) and closer to the beam in the horizontal plane. The profilemeter records the radial distribution of the particles incident on the entry face of the secondary collimator (Fig. 5). This distribution includes the peak of channeled particles deflected into the depth of the collimator, and the nonchanneled multiply scattered particles peaked at the edge of the collimator. The efficiency figures as measured on the extraction set-up were reproduced on the collimation set-up where the intensity of the channeled beam is obtained by integration of the peak in the profile.

The collimation experiment was repeated at the injection plateau of the U-70, with 1.3 GeV protons, on the same collimation set-up with the same crystal. As one can see in Figure 5 the channeling effect is still quite profound although the energy was lowered by two orders of magnitude.

### IHEP: Crystal collimation at a ramping energy

In a more recent IHEP experiment the same crystal collimation set-up was tested in a broad energy range made available in the main ring of U-70 accelerator. Earlier, the experiment was performed at the top energy flattop, 70 GeV, and at the injection flattop, 1.3 GeV, of U-70 machine. This time the tests [41] were done at seven intermediate energies and, importantly, it was not possible to arrange a flattop for each energy. During the acceleration part of the machine cycle, on a certain moment corresponding to the energy of the test, the beam was dumped in a short time onto the crystal.

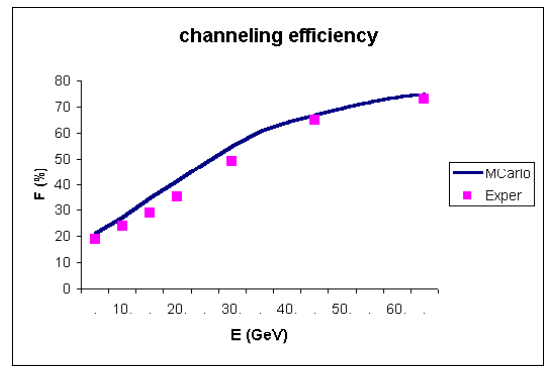

Figure 6. Crystal collimation efficiency (channeled particles ratio to the entire beam dump) as measured and as expected. The case of ramping energy in U-70.

These measurements are summarized in Figure 6 showing the ratio of the channeled particles to the entire beam dump (the *crystal collimation efficiency*) as measured and as predicted by Monte Carlo simulation.

Figure 7 shows the examples of the radial beam profile observed at the entry face of the collimator with crystal working as primary scraper, at 12 GeV and at 45 GeV. One can see that the same crystal shows efficient work from injection through the ramping up to the top energy.

The background downstream of the collimator has been measured in IHEP U70 on two locations with crystal working as a primary scraper. This background plotted versus crystal alignment in Figure 8 drops by factor of two when crystal is aligned. This experiment [42,43] has been the world first demonstration of crystal collimation. The factor of two was gained due to channeling with 50% efficiency.

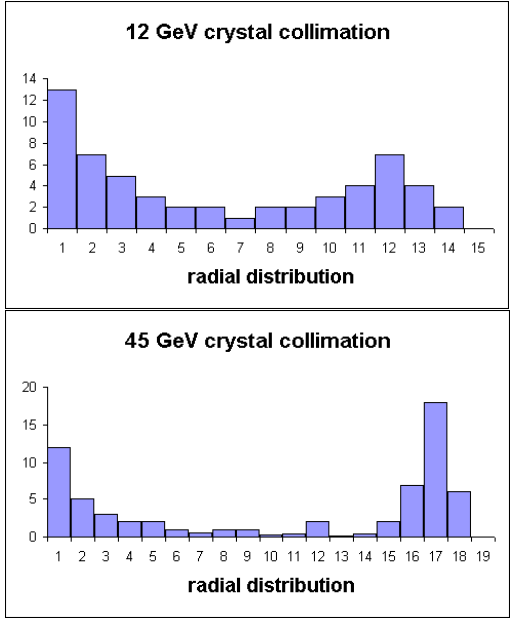

Figure 7. The radial beam profile observed at the entry face of the collimator with crystal working as primary scraper, at 12 GeV (left) and at 45 GeV (right).

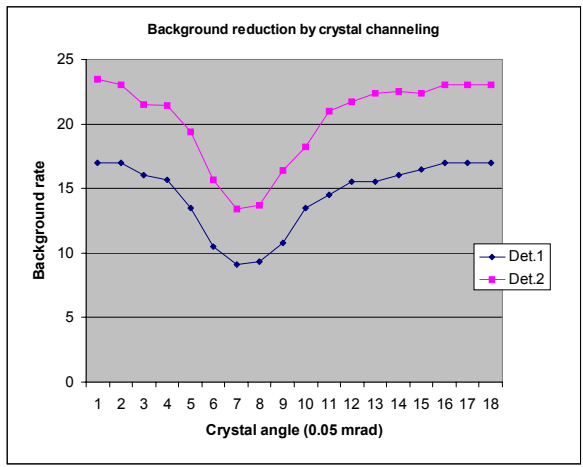

Figure 8. The background observed in IHEP U70 downstream of the collimator on two locations (Det. 1, 2) with crystal working as primary scraper, plotted versus crystal alignment.

## IHEP: Crystal collimation at high intensity

Other important issues to be addressed for a practical application of crystal-assisted extraction and collimation are thermal shock, radiation damage and crystal lifetime. In typical collimation and extraction tests at IHEP U-70, crystal channeled $\sim10^{12}$ protons (up to $3\cdot10^{12}$ in some runs) in a spill of 0.5-1 s duration.

Let us illustrate it in the following way. Suppose, all the LHC store of $3\cdot10^{14}$ protons is dumped on our single crystal in a matter of 0.2 hour [44-46]. This makes a beam of $4\cdot10^{11}$ proton/s incident on the crystal face. In IHEP, this is just routine work for a crystal, practiced every day.

One of the crystals (5 mm long) located upstream of the U-70 cleaning area was exposed for several minutes to even higher radiation flux of 70 GeV protons [5]. It received $\sim10^{14}$ proton hits per spill of 50 ms, with a repetition period of 9.6 s. Although it was impossible to characterise the crystal efficiency in such a short time, the channeling properties after the exposure of the crystal were tested in an external beam line. The deflected beam observed with photo-emulsion (Figure 9) was perfectly normal, without breaks nor significant tails eventually produced by dechanneled particles. This is a good indication of the absence of thermal and radiation damages.

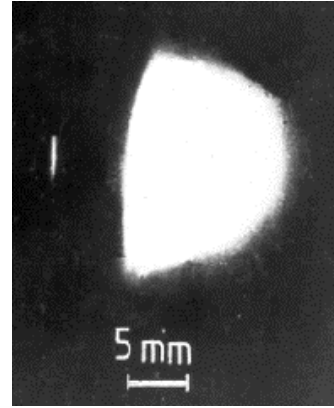

Figure 9. Photograph of the deflected (left) and incident (right) beams as seen downstream of the crystal. Prior to the test, the crystal was exposed in the ring to 50-ms pulses of very intense beam ($\sim10^{14}$ proton hits per pulse). No damage of crystal was seen in the test, after this extreme exposure.

Let us translate it to the LHC case. One bunch of the LHC is $1.1\cdot10^{11}$ protons. The IHEP crystal survives an instant dump of 1000 bunches of the LHC. The LHC collimation system is required to survive a hit of 20 bunches [44-46], so the crystal conforms to it with a great safety margin. As for the lifetime of a crystal, the CERN experiment [18] with 450 GeV protons showed that at the achieved irradiation of $5\cdot10^{20}$ proton/cm$^2$ the crystal lost only 30% of its deflection efficiency, which means about 100 years lifetime in the intense beam of NA48 experiment. One of the IHEP crystals served in the vacuum chamber of U-70 over 10 years, from 1989 to 1999, delivering beam to particle physicists, until a new crystal replaced it (in order to reduce the size of the channeled beam).

## RHIC EXPERIMENT ON CRYSTAL COLLIMATION

Another experiment on crystal collimation has been in progress at the Relativistic Heavy Ion Collider [47-51]. The yellow ring of the RHIC has a bent crystal collimator of the same type as used in earlier IHEP experiments [34], 5 mm along the beam. By properly aligning the crystal to the beam halo, particles entering the crystal are deflected away from the beam and intercepted downstream in a copper scraper. The purpose of a bent crystal is to improve the collimation efficiency as compared to a scraper alone.

Beam losses were recorded by the PIN diodes, hodoscope, and beam loss monitors. Signals from the RHIC experiments were also logged to monitor their background rates. Fig. 10 shows a typical angular scan from the 2003 RHIC run with gold ions. The same Figure 10 shows the predicted angular scan. The simulation is done with the measured machine optics. The two angular curves, measured and predicted, are in reasonable agreement.

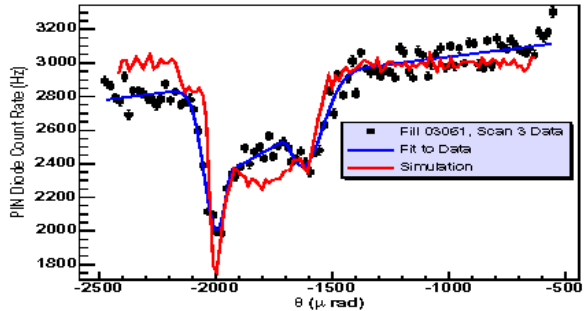

Figure 10. RHIC: Crystal collimator efficiency as a function of crystal alignment, for gold ions. Measured data and simulation (CATCH).

The efficiency is defined as maximum depth of the large dip divided by the background rate. For the 2003 RHIC run, the theory predicted the efficiency of 32%, and averaging over the data for this run gives the measured efficiency of 26%.

The modest figure of efficiency ~30%, both in theory and experiment, is attributed to the high angular spread of the beam that hits the crystal face as set by machine optics. It is worth to compare this figure of efficiency for gold ions at RHIC to the 40% efficiency achieved with similar crystal for protons at IHEP in 1998 [34]. It is also worth to notice that the crystal extraction efficiency observed at CERN SPS with lead ions was 4 to 11% with a long (40 mm) crystal of silicon [52]. The RHIC study was demonstration of world first crystal collimation for heavy ions, with efficiency record high for heavy ions.

## TEVATRON SIMULATIONS ON CRYSTAL COLLIMATION

A possibility to improve the Tevatron beam halo scraping using a bent channeling crystal instead of a thin scattering target as a primary collimator was studied at Fermilab [53]. In order to evaluate the efficiency of the collimation system, realistic simulations have been performed using the CATCH, STRUCT and MARS Monte Carlo codes.

It was shown that the scraping efficiency can be increased by one order of magnitude. As a result, the beam-related backgrounds in the CDF and D0 collider detectors can be reduced by a factor of 7 to 14. Calculated results on the system performance taking into account the thickness of near-surface amorphous layer of the crystal are presented in Table 2.

Table 2 The Tevatron: Halo hit rates at the D0 and CDF Roman pots and nuclear interaction rates $N$ (in $10^4$ p/s) in the primary scraper (target or crystal). Simulation [6]. Ten-fold improvement is expected from a crystal scraper.

|     | with target | with crystal | | |
| --- | --- | --- | --- | --- |
|     |     | amorphous layer thickness | | |
|     |     | 10 μm | 5 μm | 2 μm |
| DØ  | 11.5 | 1.35 | 1.60 | 1.15 |
| CDF | 43.6 | 5.40 | 3.20 | 3.43 |
| $N$ | 270 | 82.4 | 70.6 | 50.3 |

Two cases have been compared:
1. The Tevatron RUN-II collimation system with all secondary collimators in design positions, but only one (D17h) horizontal primary collimator in working position. This primary collimator intercepts large amplitude protons and protons with positive momentum deviations.
2. The same collimation scheme, but silicon bent crystal is used instead of primary collimator.

## SIMULATIONS FOR THE LHC

We evaluated the potential effect of crystal collimation in the LHC using the same computer model [10-12] already validated with the IHEP, CERN SPS, Tevatron, and RHIC experiments on channeling. Simulations were done in the LHC both at the collision energy of 7 TeV and at the injection energy of 450 GeV for a nominal beam emittance of 3.75 μm (at 1 σ, defined as r.m.s.). In the model, a bent crystal was positioned as a primary element at a horizontal coordinate of 6σ in the halo of the LHC beam, on one of the locations presently chosen for amorphous primary elements of the LHC collimation system design [46]. The LHC lattice functions were taken corresponding to this position: $\alpha_x$=1.782 and $\beta_x$ =119 m in the horizontal plane, and $\alpha_y$ = −2.02 and $\beta_y$ =143.6 in the vertical plane. With the previous parameters the horizontal value of σ for the beam is of 0.95 mm.

In this study we did not investigate the effect on channeling efficiency of the lattice functions at the location of the crystal. Such an effect may be rather important as shown by recent RHIC results [47-51], where a good agreement was found between the prediction (based on the same model used here for the LHC) and the measurement only when the measured lattice functions were also taken into account.

Instead, we varied crystal parameters such as the size, bending, alignment angle, material, and the quality of the surface. We observed the efficiency of channeling, i.e., the number of the particles deflected at the full bending angle of the crystal, taking into account many turns in the LHC ring and multiple encounters with the crystal.

### Optimal size of the crystal

On the first encounter, the halo particles were assumed to enter in the crystal face within ≤1 μm from the edge. In addition, the first 1-μm thick near-surface layer of the crystal was assumed to be amorphous. This means that, during the first encounter, the particles were not channeled at all: they were just scattered after traversing the full crystal length. This is a rather conservative approach. In principle, it is possible to make a crystal face of a much better accuracy, with less that ~0.01 μm thick amorphous layer (an oxidized layer thickness) [54]. In this case, most of the incident halo particles would be channeled already on the very first encounter.

Fig. 11 shows the computed channeling efficiency as a function of the crystal length along the LHC beam for two cases: at flattop (7 TeV) and at injection (450 GeV). Two bending angles were used. The optimal size of the silicon crystal is about 10 mm for 0.2-mrad bending, 5 mm for 0.1 mrad, and 3 mm for 0.05 mrad. High efficiency of channeling can be obtained with the same (optimized) crystal both at 7 TeV and at 450 GeV. The efficiency is expected to be 90-94% in the case of crystal bending angle of 0.05-0.1 mrad.

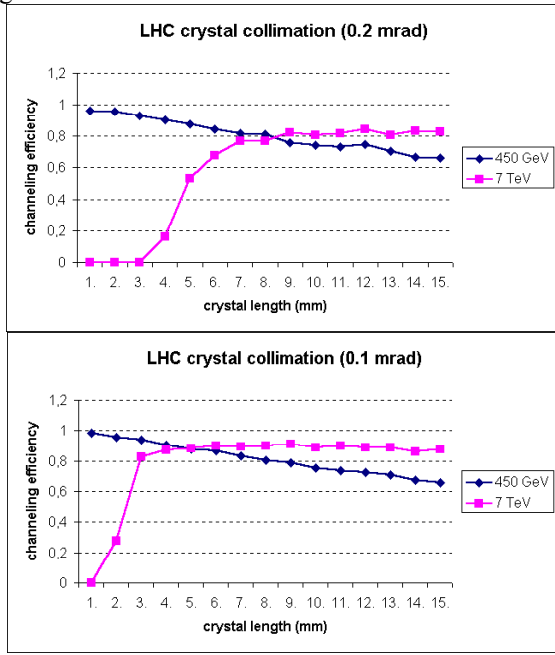

Figure 11. Channeling efficiency as a function of the crystal length along the LHC beam, shown for two cases: at flattop and at injection. The top figure is for the crystal bending of 0.2 mrad, the bottom one for 0.1 mrad.

## Estimation of halo reduction

Different bending angles were examined (finding every time the optimal size for the crystal) and the channeling efficiency computed. Fig. 12 (top) shows the channeling efficiency, F, as a function of the crystal-bending angle. Fig. 12 (bottom) shows the same data plotted as a "background reduction factor" $1/(1-F)$, namely the factor that halo intensity is expected to be reduced. If all channeled particles were fully intercepted by the secondary collimator, then only non-channeled particles should contribute to the background in the accelerator.

It should be said that all the range of crystal deflector size assumed in Fig. 12 is already realised and tested by IHEP in 70 GeV beam. This technique is available and well established.

## Choice of the crystal material: low and high Z

The optics of traditional (amorphous) collimation at accelerators and technical considerations may require primary scrapers of different material (atomic number Z). As the technique of bent crystal channeling is developed also with other materials, e.g. germanium (Z=32) [55,56] and diamond (Z=6) [57,58], we continued our studies with other crystals.

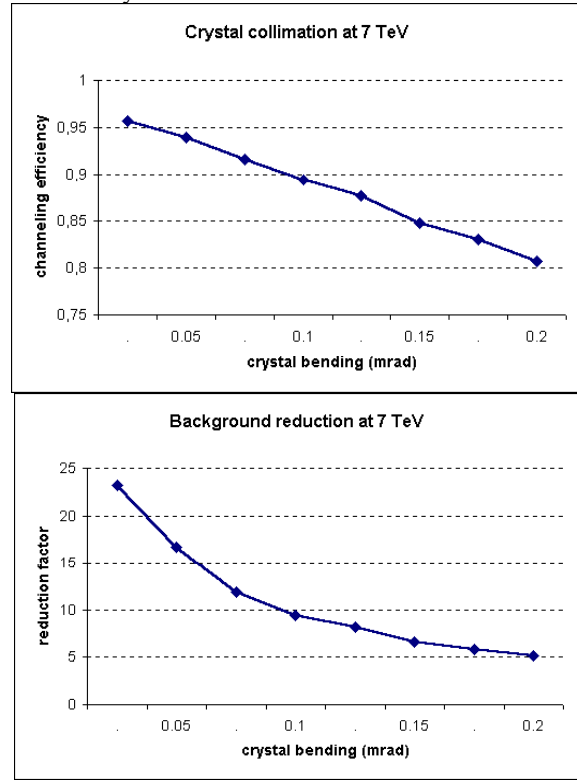

Figure 12. The top plot is the channeling efficiency $F$ vs crystal bending angle. The bottom plot is the same data plotted as an LHC background reduction factor $1/(1-F)$. For Si(110) crystal with a rough (1 micron) surface.

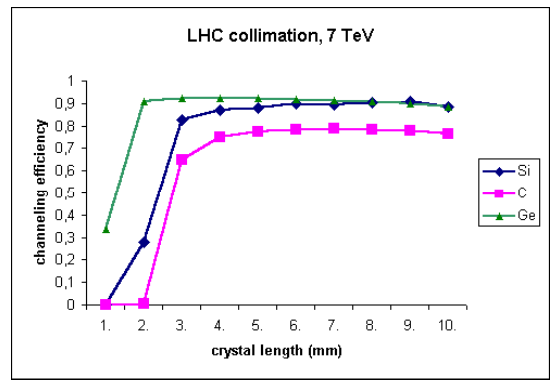

Fig. 13 Channeling efficiency at 7 TeV as a function of the crystal length along the LHC beam, shown for different crystals: Silicon (Z=14), Diamond (Z=6), and Germanium (Z=32).

Fig. 13 gives the channeling efficiency for different crystals. Crystal plane (110) and bending 0.1 mrad were used in each case. We see that comparable efficiencies can be obtained in all these cases. All these crystals, from diamond to germanium, can serve as an LHC primary scraper. Another interesting (although futuristic) possibility might be the use of nanostructured material [59-62], single-wall and multi-wall nanotubes which appear to be potentially useful as material for channeling collimation [63]: see Figure 14.

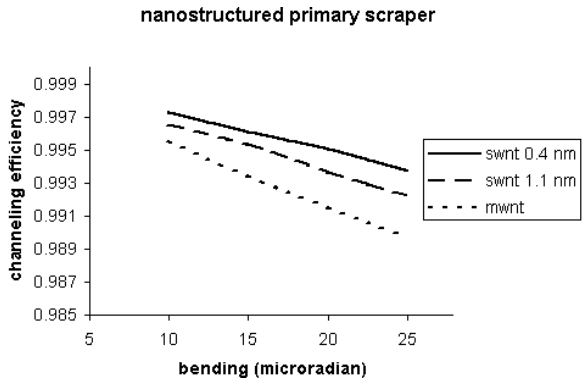

Figure 14. The overall (multi-pass) channeling efficiency $F$ of nano-structured scraper shown as a function of bending angle for SWNT (0.4 and 1.1 nm diameter) and MWNT at 7 TeV.

*Alignment precision*

For an efficient operation, the crystal must be oriented parallel to the envelope of the circulating beam. Fig. 15 shows the computed channeling efficiency as a function of the crystal orientation angle w.r.t. the incident particles in the LHC. The orientation curve has FWHM of about 7 µrad at top energy. It is worth mentioning that the same problem of alignment accuracy is present in a conventional collimation system. For instance, in the baseline collimation system for LHC the primary collimator should be aligned with an accuracy of 20 µrad.

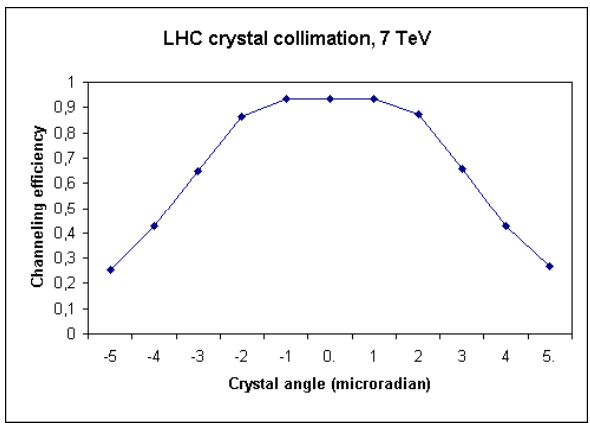

Figure 15. Channeling efficiency as a function of the crystal orientation angle w.r.t. the beam envelope. The orientation curve has FWHM ≈ 7 µrad at top energy.

*Crystal surface specifications*

The traditional approach to the LHC collimation system requires that the primary scraper should have flat surface with the accuracy (non-flatness) of 10 micron. The above simulations assumed crystal surface roughness of 1 µm. We have studied the effect of surface roughness to channeling efficiency at the LHC, modeling the near-surface non-channeling layer ("septum width") as if it was amorphous.

Fig. 16 shows channeling efficiency as a function of the crystal surface roughness. Even a crystal with a relatively rough surface—irregular by 10 µm—shows high efficiency. The reason of such behavior relies on the multi-turn scheme for extraction. By making the surface perfect to a better level than 1 µm, the computed efficiency will exceed 97%.

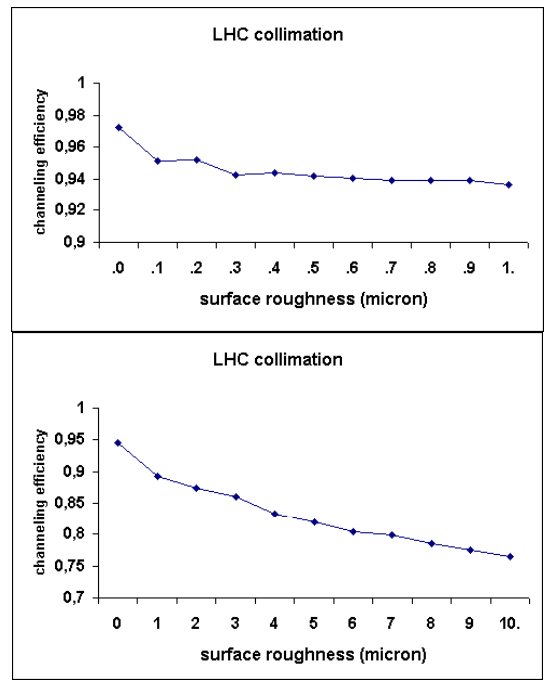

Figure 16. Channeling efficiency as a function of the crystal surface roughness ("septum width"). Two ranges shown: 0-1 µm (left) and 0-10 µm (right).

This dependence can be compared to the studies of crystal extraction efficiency in simulations [30-32,64] for the SPS as a function of crystal surface roughness ("septum width"), Figure 17.

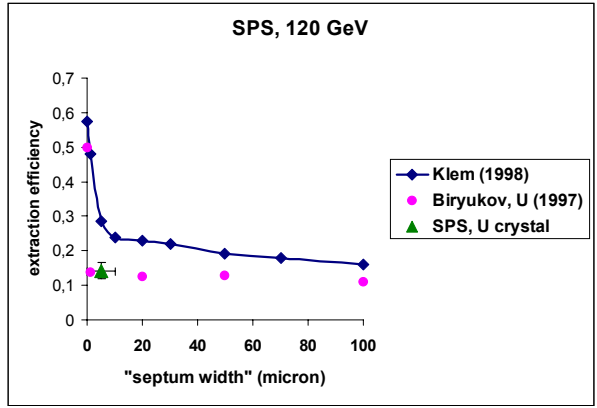

Figure 17. Crystal extraction efficiency (absolute value) versus crystal surface roughness ("septum width") at the SPS. Two models (1997, 1998) and SPS data.

## POSSIBLE DESIGNS FOR CRYSTAL SCRAPER: *S-TYPE VS O-TYPE*

At present, there are two options of what kind of crystal to try for the collimation experiments. One option is O-type crystals heavily used at IHEP and tested at RHIC, another option is Strip-type crystal (again, Strip types were heavily tested and exploited at IHEP). These two particular crystals, O-type and S-type, may have about the same size (5 mm) along the beam and bending angle (e.g. order of 0.5 mrad), however they strongly differ in design. The O-type has straight sections ("legs") while S-type is bent uniformly without straight parts. Apart from the choice from the crystal deflectors already available, it is important to understand what kind of crystal should be optimal for future studies.

Here we made a preliminary Monte Carlo simulation to compare the two crystal designs, O-type and S-type. Both crystals had 5 mm size along the beam of 900 GeV protons, and bending of 0.5 mrad. The O-type had straight parts 1+1 mm; therefore only 3 mm were bent. These parameters are close to what we have in some of the first crystals prepared for next run of experiments at FNAL.

This is not as detailed simulation as the one published at PAC 1999 [53], because the Tevatron environment relevant for collimation studies was not considered in full. However this simulation is sufficient to make predictions for the relative performance of the two types of crystal. We assumed that all channeled particles were intercepted safely by an absorber, therefore only nonchanneled particles contributed to the background in the Tevatron.

Figure 18 shows, in relative units, the background produced in the Tevatron ring by the two types of crystal as a function of crystal orientation. The orientation angular range where crystal channels best is about 8 microradian full width. Under this orientation, S-type crystal produces background lower than O-type crystal does, by a factor of 1.60±0.05 because of the channeling with higher efficiency.

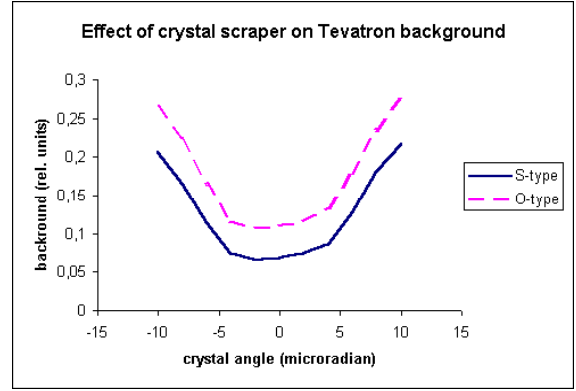

Figure 18. Comparison of two designs for crystal scraper.

The conclusion of this comparative study: Strip type crystal deflector produces essentially lower background in Tevatron than O-type crystal does, by a factor of 1.60±0.05 according to tentative Monte Carlo simulations for crystal channeling scraping in Tevatron, for other crystal parameters like size and bending being fixed. This conclusion agrees with the IHEP experimental practice where typical *inefficiency* achieved with S-type crystal has been 0.15 (i.e. efficiency of 85%) while O-type crystals have shown inefficiency of 0.35-0.60 (i.e. efficiency of 40-65%), i.e. performance factor of 2-4 weaker than that of S-type crystal deflectors.

Tevatron tests of crystal collimation will provide the best opportunity for validation of the technique; however, lots of new information can be gained from the tests in the SPS. First preliminary Monte Carlo studies for the SPS predict that S-type crystal could show efficiency as high as 95% in collimation experiment if crystal size is chosen close to the one used at IHEP, 2 mm along the beam. Here we assumed that crystal collimation tests are run at 270 GeV in straight section 5 of the SPS with bending angle of 0.2 mrad.

Simulations and experiment have identified strip-type crystals as the choice for collimation in the LHC, SPS, Tevatron. It is advised to install a strip-type crystal as a scraper for the first collimation tests.

Further Monte Carlo studies are advisable with a realistic account of collimation settings and with a study of other crystal parameters like size and bending angle in order to find the best crystal option for the planned experiments on crystal collimation, and to have detailed theoretical predictions.

## CONCLUSION

Crystal would be very efficient in the LHC environment. The expected efficiency figure, ~90%, is already experimentally demonstrated at IHEP and confirmed by simulations for the Tevatron. This will make the LHC 10 times (up to ~40 times) cleaner. Monte Carlo model successfully predicts the crystal work in the circulating beam, as demonstrated recently in crystal collimation experiments at IHEP and RHIC, and in crystal extraction experiments at up to 900 GeV (the Tevatron).

Crystal works efficiently at very high intensities (~$10^{12}$), actually much higher than the LHC requires, with a lifetime of many years. Crystal survives the abnormal dump of the LHC beam with ~100-fold safety margin (i.e. survives the instant dump of 1000 LHC bunches or ~$10^{14}$ protons) as demonstrated experimentally at 70 GeV.

The same crystal scraper works efficiently over full energy range, from injection through ramping up to top energy, as demonstrated experimentally at IHEP from 1 through 70 GeV and as seen in simulations for the LHC. Bent crystals of low-Z and high-Z material are available, e.g. diamond and germanium, and they demonstrate the efficiency similar to that of silicon. Even when a crystal is misaligned, nonchanneling, it still works as an amorphous scatterer so the collimation system returns to its traditional scheme. This makes it safe.

## ACKNOWLEDGEMENTS

We acknowledge the support of the European Community-Research Infrastructure Activity under the FP6 "Structuring the European Research Area" programme (CARE, contract number RII3-CT-2003-506395). This work was supported by INTAS-CERN grant 03-52-6155.